\documentclass[conference]{IEEEtran}
\IEEEoverridecommandlockouts

\usepackage{cite}
\usepackage{amsmath,amssymb,amsfonts}
\usepackage{algorithm}
\usepackage{algpseudocode}
\usepackage{url} 
\usepackage{hyperref} 
\usepackage{graphicx}
\usepackage{textcomp}
\usepackage{xcolor}
\usepackage{algorithm}
\usepackage{booktabs}
\usepackage{siunitx} % For number alignment
\usepackage{microtype}
\usepackage{booktabs}  % for \toprule, \midrule, etc.
\def\BibTeX{{\rm B\kern-.05em{\sc i\kern-.025em b}\kern-.08em
    T\kern-.1667em\lower.7ex\hbox{E}\kern-.125emX}}
\begin{document}
\title{Leveraging LLMs for Dynamic IoT Systems Generation through Mixed-Initiative Interaction}

\author{\IEEEauthorblockN{Bassam Adnan\textsuperscript{\dag}}
\IEEEauthorblockA{\textit{IIIT Hyderabad, India}\\
bassam.adnan@research.iiit.ac.in}
\and
\IEEEauthorblockN{Sathvika Miryala\textsuperscript{\dag}}
\IEEEauthorblockA{\textit{IIIT Hyderabad, India}\\
miryala.sathvika@research.iiit.ac.in}
\and
\IEEEauthorblockN{Aneesh Sambu\textsuperscript{\dag}}
\IEEEauthorblockA{\textit{IIIT Hyderabad, India}\\
sambu.aneesh@research.iiit.ac.in}
 \and
\IEEEauthorblockN{Karthik Vaidhyanathan}
\IEEEauthorblockA{\textit{IIIT Hyderabad, India}\\
karthik.vaidhyanathan@iiit.ac.in}
\and
\IEEEauthorblockN{Martina De Sanctis}
\IEEEauthorblockA{\textit{GSSI, L’Aquila, Italy}\\
martina.desanctis@gssi.it}
\and
\IEEEauthorblockN{Romina Spalazzese}
\IEEEauthorblockA{\textit{Malmö University, Sweden}\\
romina.spalazzese@mau.se}
}
\maketitle

\renewcommand{\thefootnote}{\dag} %  dagger for this footnote
\footnotemark
\footnotetext{These authors contributed equally to this work.}
\renewcommand{\thefootnote}{\arabic{footnote}} % Restore footnote numbering
\setcounter{footnote}{0}

\begin{abstract}
IoT systems face significant challenges in adapting to user needs, which are often under-specified and evolve with changing environmental contexts. To address these complexities, users should be able to explore possibilities, while IoT systems must learn and support users in the process of providing proper services, e.g., to serve novel experiences. The IoT-Together paradigm aims to meet this demand through the Mixed-Initiative Interaction (MII) paradigm that facilitates a collaborative synergy between users and IoT systems, enabling the co-creation of intelligent and adaptive solutions that are precisely aligned with user-defined goals. This work advances IoT-Together by integrating Large Language Models (LLMs) into its architecture. Our approach enables intelligent goal interpretation through a multi-pass dialogue framework and dynamic service generation at runtime according to user needs. To demonstrate the efficacy of our methodology, we design and implement the system in the context of a smart city tourism case study. We evaluate the system's performance using agent-based simulation and user studies. Results indicate efficient and accurate service identification and high adaptation quality. The empirical evidence indicates that the integration of Large Language Models (LLMs) into IoT architectures can significantly enhance the architectural adaptability of the system while ensuring real-world usability.

\end{abstract}

\begin{IEEEkeywords}
LLM, Self-Adaptation, Software Architecture, Service Generation, Dynamic Application Generation, IoT-Together Paradigm
\end{IEEEkeywords}

\section{Introduction}

The IoT-Together \cite{spalazzese2023shaping}, a Mixed-Initiative Paradigm \cite{horvitz1999principles} promotes collaboration between users and intelligent IoT systems, where both actively shape and co-generate IoT solutions. This paradigm helps users explore suitable options for their needs and preferences, especially in dynamic, underspecified IoT systems. For example, in a Smart City scenario, a tourist arrives without a plan and, through collaboration with the system, identifies an itinerary that matches their preferences. The paradigm encompasses several key aspects. First, it recognizes that user goals may be unclear or evolve over time, allowing users to express their goals and preferences dynamically, such as finding activities in a smart city or indicating a preference for walking. Second, it establishes an initial interaction point, typically via smartphone or totem, where users can define their needs, setting the stage for ongoing collaboration. Third, the paradigm supports dynamic, real-time IoT system generation that adapts to user's evolving needs. Finally, it emphasizes the need for designing systems that can clearly explain actions and recommendations, helping users make informed decisions without feeling overwhelmed.

\begin{figure*}[ht]
    \centering
    \includegraphics[width=0.80\textwidth]{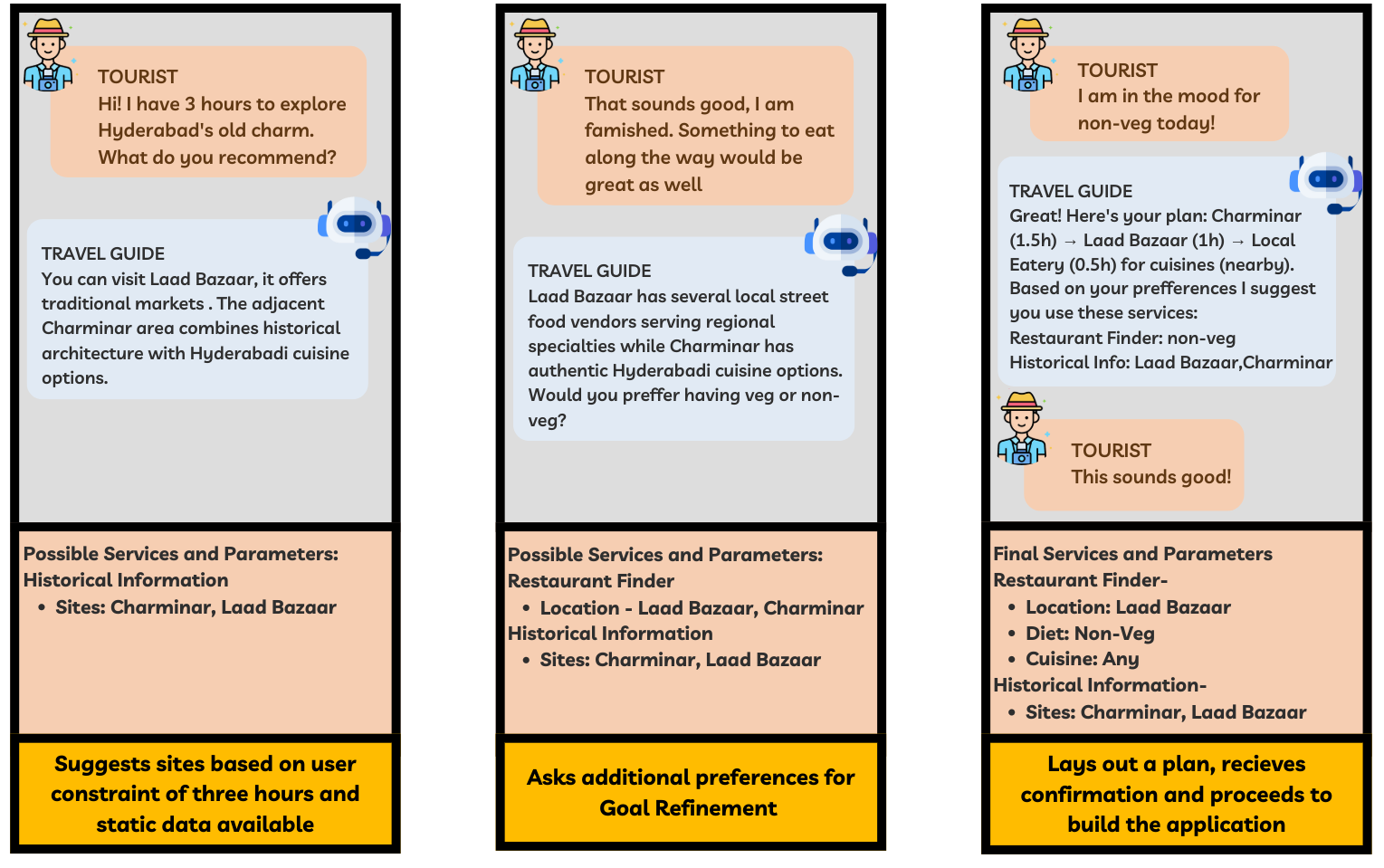} 
    \label{fig:initial-interaction-diagram}
    \caption{Three-Pass Dialogue Flow: Progressive Identification of User Goals and Service Parameters enabling Goal-Driven Architecture}

\end{figure*}

While the IoT-Together Paradigm offers a conceptual framework, realizing key components such as dynamic goal understanding and system generation presents significant challenges. Large Language Models (LLMs) offer promising capabilities in addressing these challenges, particularly due to their adept natural language understanding and structured knowledge extraction abilities, which are essential for implementing the paradigm's core aspects of natural interaction and facilitate dynamic system adaptation\cite{he2024doespromptformattingimpact}. Also, existing IoT frameworks often rely on pre-defined service catalogs, limiting their ability to adapt dynamically when user goals require services not already implemented\cite{7479556}. To this end, we propose an LLM-powered architecture that enables dynamic goal derivation from user interactions and runtime service generation based on available IoT definitions and data schemas. This ensures that new services are only generated when corresponding sensor data and service descriptions are present, grounding the system's adaptability in its existing context. Fig.~\ref{fig:initial-interaction-diagram} demonstrates our realized approach, showing how the system engages in structured dialogue with users to understand their needs, processes these inputs through LLM-powered components, and generates appropriate IoT solutions—a concrete implementation of the interaction framework that was conceptually proposed in the original paradigm.

Our architecture implements these capabilities through dynamic goal extraction from natural language user queries, real-time service composition based on available IoT definitions, and adaptive service generation that aligns with both user preferences and system constraints. To validate our approach, we present a smart city case study centered in Hyderabad. The evaluation employs a multi-agent simulation framework and a user study. We assess system performance through multiple dimensions: (1)~service identification accuracy (2)~code generation quality and (3)~system efficiency through token consumption and latency measurements. The evaluation demonstrates the system's capability to identify user needs along with the generation of appropriate services based on their needs while maintaining efficiency in both service generation and overall application composition.

\section{Motivating  Case Study}
To motivate our work and demonstrate the realization of the IoT-Together paradigm, we present a smart city system developed for the city of Hyderabad, India. The system has been implemented at the Smart City Living Lab of The International Institute of Information Technology Hyderabad, building upon the collaborative principles established in the IoT-Together paradigm~\footnote{\url{https://smartcityresearch.iiit.ac.in}}. The existing smart city infrastructure comprises a network of IoT sensors that monitor crowd density in specific regions, air quality at public spaces, and water quality at heritage sites. Currently, visitors to the city need to use multiple separate applications to access information about restaurant availability, cultural heritage site bookings, and other services. This fragmentation becomes particularly challenging when visitors have uncertain or evolving objectives - a key scenario that the IoT-Together paradigm aims to address. Planning activities requires complex decision-making that considers user preferences, time constraints, and environmental conditions.

We provide a system consisting of a web application that integrates with the city's IoT infrastructure, collecting contextual data from various urban locations. The sensor network provides real-time data on crowd density, air quality, and water quality measurements at public places such as forts, museums, historical sites, restaurants, hospitals. Built upon this sensor infrastructure are several services: event notifications, historical site information systems, crowd monitoring, air quality assessment tools, exhibition tracking, restaurant recommendation engines, ticket purchasing platforms, travel planning tools, and water quality monitoring services. These services highlight the opportunity for applying IoT-Together's collaborative approach in helping users to achieving their goals.

\noindent In the following section to explain our approach, we consider an instance of the above case study, where a tourist visiting the smart city has a limited time window of three hours to explore the historical and cultural attractions.

\section{Approach}
\begin{figure*}[ht]
    \centering
    \includegraphics[width=1.4\textwidth, height=10cm, keepaspectratio]{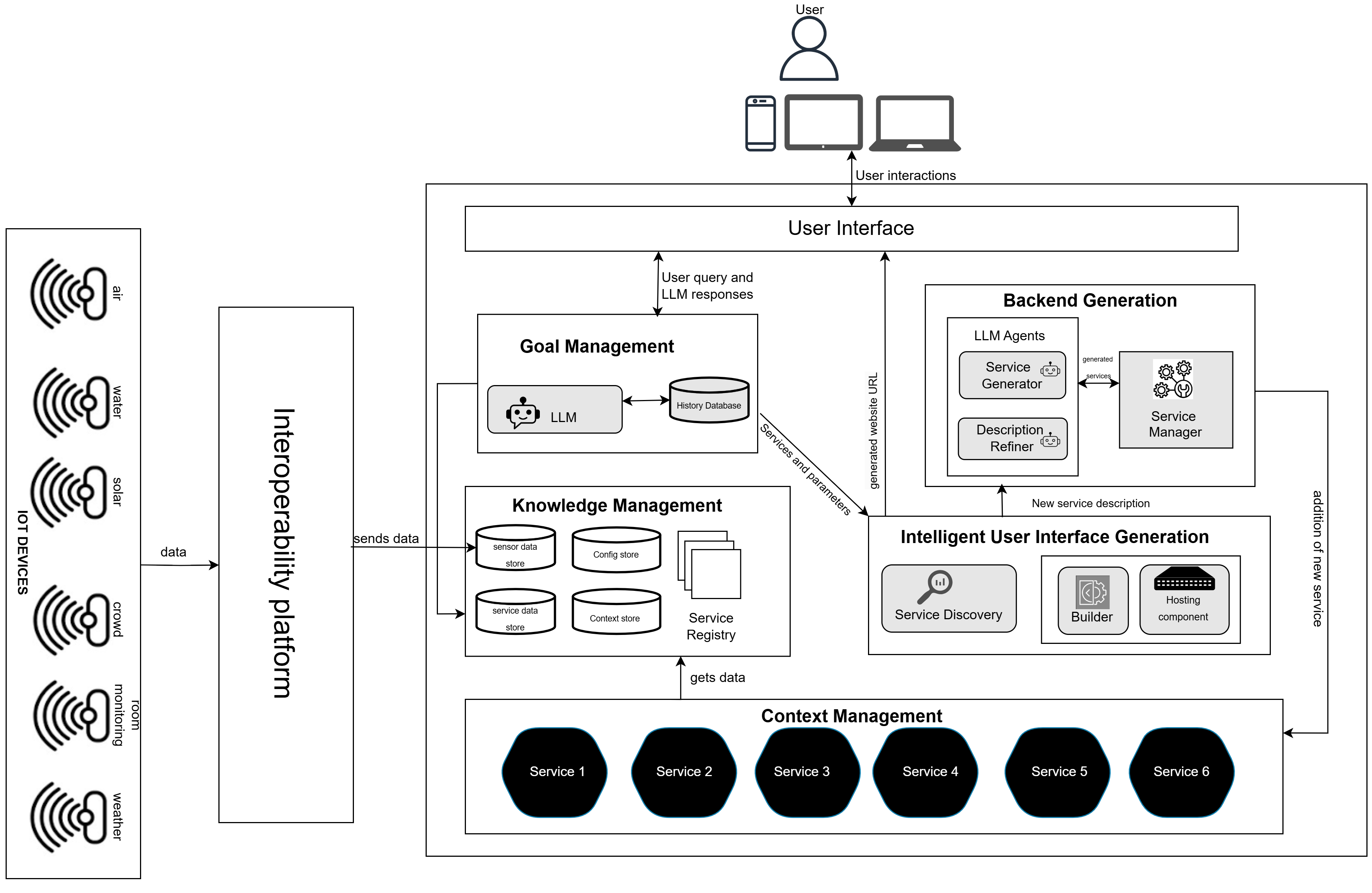}
    \caption{High-Level Architecture: System Components and Their Interactions}
    \label{fig:high-level-architecture-overview}
\end{figure*}

Figure \ref{fig:high-level-architecture-overview} presents the proposed system architecture that implements the \textit{IoT-Together paradigm}. The newly proposed components, distinguished by grey shading in the figure, form an integral part of the overall system design. While the \textit{learning management} component remains unimplemented in the current version, it has been identified as a key area for future development. The system facilitates dynamic application reconfiguration based on user goals through comprehensive integration of IoT environmental data and system-level information. The system architecture is designed in a way such that, it supports dynamic evolvability through the generation and integration of new services at run-time in accordance with user goals.

The system adopts key components from the IoT-Together paradigm, including \textit{Goal Management}, \textit{Knowledge Management}, \textit{Context Management}, \textit{Intelligent User Interface (IUI) Generation}, and \textit{Backend Generation}. User interacts with the system using a device (Smart Phone/Laptop/Tablet) through the user interface to enter the query. This query is then passed to the Goal Management, which identifies the set of services that satisfy user goals. The Goal Management uses the LLM to determine the services within the system that can satisfy the user goals or the services that need to be generated. In this phase, the user and system collaboratively (through back-and-forth conversational exchanges) identify the set of parameters and the services required to achieve the user goal.
After the identification of the required services, they are transferred to the Intelligent User Interface Generator to build the application. It locates the services in the Knowledge Management; otherwise, a description of the new service is forwarded to the Backend Generation component, which generates the service, thus extending the architecture runtime according to the user needs. The generated service is sent back to the Intelligent User Interface Generator. The builder within the component integrates these services using a pre-defined template and then creates the website. Intelligent User Interface Generator provides the URL to access this website created. Fig-3 provides the sequence diagram for the process of generation of the application to the user. Meanwhile, the sensors of the IoT environment transmits data periodically to Knowledge Management, which maintains a persistent data repository. This accumulated data is subsequently utilized by the system during the creation of new services and translating the user needs to the goals using the Goal Management.
\\
The User Interface serves as the interaction medium between users and the system, managing conversational exchanges during service identification and preference elicitation. Users submit their initial queries through the interface, and the system engages in a structured conversation to clarify requirements. We represent user queries by $Q_\text{user}$ and system responses by $R_\text{sys}$. For the running example, $Q_\text{user} =$ ``I have 3 hours to explore Hyderabad's old charm" and the $R_\text{sys}$ provides relevant information along with follow-up queries for goal formulation.
Upon successful completion of this dialogue, the interface presents the user with a URL to access the generated web application that implements their specified needs.\\
In the subsequent, following the Fig- \ref{fig:high-level-architecture-overview} we introduce core components and the interactions between them in detail.

\subsection{\textbf{Goal Management}}
User Interface forwards user queries to the Goal Management. The LLM in the Goal Management interprets these queries based on three elements: the summary of previous exchanges, the environmental context from Knowledge Management, and the current state of system services. These conversations are modeled to extract maximum information about user goals and preferences. The LLM engages in an iterative dialogue with users to progressively refine the user goals until all requisite services are identified. This extraction process is carried out using prompt templates in LLMs, each designed to identify specific contextual elements at each pass: environmental parameters in the first pass, goal refinement in the second pass, and service requirements in the final pass. User preferences are then applied as parameters to customize the identified services. To enable this information extraction process, we designed a three-pass conversation structure.

\textbf{Pass 1: Contextual Awareness} 
In this pass, the LLM aggregates contextual information $C$ by combining sensor data $D_{\text{sensor}}$ (weather conditions, noise levels, time of day, day of the week, traffic conditions, and air quality), with user-specific data $D_{\text{user}}$ (current location, past activities, and stated preferences as $\text{params}$). A contextual state $X$ maintains the current interaction state, while service registry data from Knowledge Management provides information about available service capabilities. Each service is represented as below:
\[
S_i = ( \text{name}_i, \text{description}_i, \text{params}_i ), \quad \text{for} \quad i = 1, 2, \dots, n
\]
 Realization of values for few services are shown in Table ~\ref{tab:service-definitions}.
\begin{table}[ht]
\setlength{\tabcolsep}{2pt}
\caption{Service Definitions\protect\footnotemark}
\label{tab:service-definitions}
\centering
\begin{tabular}{lll}
\toprule
\textbf{Service Name} & \textbf{Description} & \textbf{Parameters} \\
\midrule
historical\_info & Provides historical and cultural & site\_name \\
 & information about monuments and sites & \\ \hline
restaurant\_finder & Recommends restaurants based on & location, cuisine, \\
 & cuisine preferences and location & diet \\ \hline
crowd\_monitor & Tracks and reports real-time crowd & location, time \\
 & density at various locations & \\ 
\bottomrule
\end{tabular}
\end{table}
\footnotetext{The values presented are illustrative and not representative of entire data. Definitions may also include associated schema, endpoints, and additional metadata relevant to the service.}
Together these services are used as 
\[
D_{\text{services}} = \{ S_1, S_2, \dots, S_n \}
\]
The complete context is defined as \[C = \{D_{\text{sensor}}, D_{\text{user}}, D_{\text{services}}, X\}\]

For example, when a user visits the city with time constraints and specific interests, such as local cuisine, nature activities, or historical sites, their preferences are stored as contextual information.

In the running example, the system identifies the user as a history enthusiast along with the time constraints and captures it via prompt templates.
\[
D_{\text{user}} = \{
\text{current location}, \text{3hrs time}, \text{history enthusiast}
\}
\]
After establishing the complete context, the next pass proceeds with goal identification. \\
\textbf{Pass 2: Goal Formulation and Refinement}
In this pass, the LLM generates Initial Goal Hypotheses based on contextual insights about the user's potential goals $G_{\text{user}}$. These hypotheses range from broad intentions (e.g., leisure activities) to concrete objectives (e.g., specific venue requirements). The system employs \textbf{Mixed-Initiative Interaction}, establishing a collaborative dialogue between the user and system for goal refinement. This refinement process is internally guided by structured prompts that evaluate hypotheses against available services and contextual constraints.

This interaction comprises two key components, each driven by specialized prompt templates:
\begin{itemize}
    \item \textbf{Proactive Suggestions}: The system suggests possible activities based on initial goal hypotheses and context $C$, using prompts that analyze alignment between user preferences and available services
    \item \textbf{Clarification Dialogues}: The system elicits specific preferences and constraints through targeted queries, using prompts designed to resolve ambiguities and gather missing information
\end{itemize}

For the running example, the Goal Management uses these structured prompts to identify historical information services as a potential match. This is represented as:
\[
S_{\text{hist}} = \{
    \text{name}: \text{historical\_info},
    \text{description},
    \text{params}: \text{site\_name}
\}
\]

where $S_{\text{hist}} \in D_{\text{services}}$ represents the identified service matching the user's interests in $D_{\text{user}}$. \\
\textbf{Pass 3: Goal Verification and Confirmation}
In this final pass, the LLM proposes a curated set of services that align with the inferred goals and contextual factors extracted from the prompt templates gathered from \textbf{Pass 1} and \textbf{Pass 2}. If the user rejects the final proposal, the process reverts to \textbf{Pass 1}.

Continuing the running example, the system first seeks confirmation for the historical information service. Upon user request for additional dining options, the system extends the service set as follows:
\[
G_{\text{users}} = \{ S_{hist},S_{restau}
\}
\]

where $S_{\text{hist}}$ and $S_{\text{restau}}$ follow from Table~\ref{tab:service-definitions} for Historical Information and Restaurant Finder services respectively, with parameters instantiated based on user preferences. In our running example, for the historical information service and restaurant finder, these parameters realize the values based on user conversations with system specifying their requirements and preferences.\\ 
\(
\theta_q = \{
site\_name = 
    \text{[Charminar} \text{, Laad Bazaar]},
\)
\(
location = 
    \text{Laad Bazaar}, 
cuisine = \text{Any},
\)
\(
diet=\text{Non-vegetarian}\}
\)
\\The system maintains a history data base $H$  along with contextual information throughout these passes which include:

\[
H = \{P_{\text{user}}, S_{\text{identified}}, C_{\text{prev}}\}
\]

where $P_{\text{user}}$ represents user preferences, $S_{\text{identified}}$ tracks identified services, and $C_{\text{prev}}$ maintains the contextual history of previous interactions.
\begin{figure*}[ht]
    \centering
    \includegraphics[width=\textwidth]{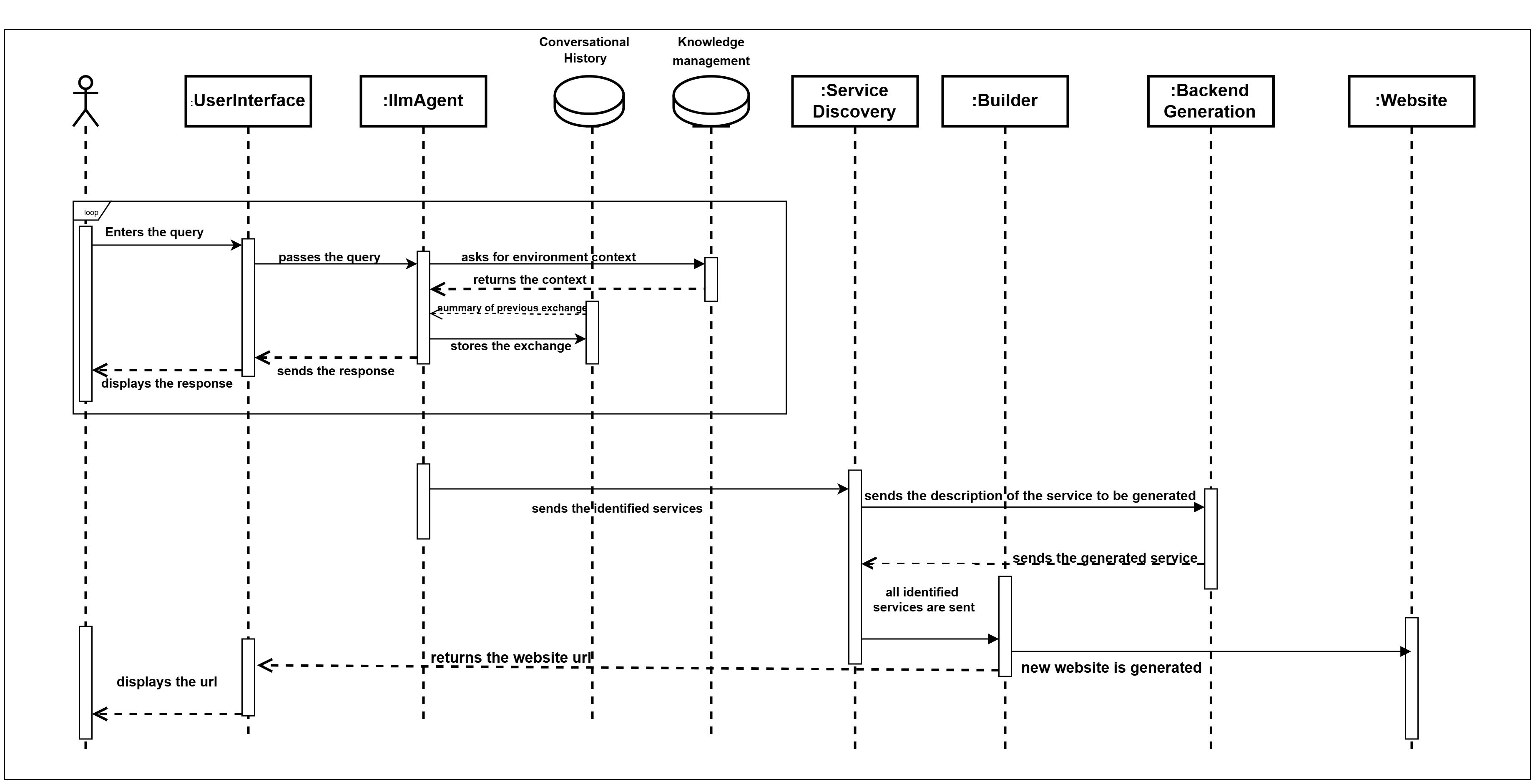}
    \label{sequence}
    \caption{Sequence diagram for creating the web application }
\end{figure*}

\subsection{\textbf{Backend Generation}}

The Backend Generation component activates when the Service Discovery identifies service requirements from the Goal Management that can be fulfilled based on available definitions but are not yet implemented in the current service set $D_{\text{services}}$. 
\begin{algorithm}
\caption{: Service Generation}
\label{alg:service_gen}
\begin{algorithmic}[1] % Ensure sequential line numbering
\State \textbf{Input:} Service requirement query $Q$, System context $S_{\text{ctx}}$
\State \textbf{Output:} Service $S$ (either matched or newly generated)
\\
\Function{ServiceGeneration}{$Q, S_{\text{ctx}}$}
    \State $D_{\text{services}} \gets S_{\text{ctx}}.services$
    
    \State // description refinement phase
    \State $service\_match \gets \text{DescriptionRefiner}(Q, D_{\text{services}})$
    \If{$service\_match \neq \emptyset$}
        \State // existing service satisfies Q
        \State \Return $service\_match$
    \EndIf
    
    \State // service generation phase
    \State $D_{\text{schema}} \gets S_{\text{ctx}}.schemas$
    \State $Q_{\text{gen}} \gets \text{RefineGenerationQuery}(Q, D_{\text{schema}})$
    \State $S_{\text{new}} \gets \text{GenerateService}(Q_{\text{gen}})$
    \State $D_{\text{services}} \gets D_{\text{services}} \cup \{S_{\text{new}}\}$
    \State $\text{UpdateServiceContext}(S_{\text{ctx}}, S_{\text{new}})$
    \State \Return $S_{\text{new}}$
\EndFunction
\end{algorithmic}
\end{algorithm}
An algorithm to achieve this is described in Service Generation (Algorithm-\ref{alg:service_gen}) with components described in Table-\ref{tab:service-gen-components}. The generation process initiates by creating a service requirement query:
\[
Q = \{
    type: \text{service\_req},
    params: \theta_q
\}
\]
where $\theta_q$ represents the required service parameters  identified through the dialogue exchange with the Goal Parser. \\

The Service Manager maintains the system context $S_{\text{ctx}}$ comprising:

\[
S_{\text{ctx}} = \{
    D_{\text{services}},
    D_{\text{schema}},
    D_{\text{config}}
\}
\]

where $D_{\text{schema}}$ represents available database schemas and $D_{\text{config}}$ contains service configurations. The description of the services assist the Description Refiner -- an LLM Agent -- to evaluate incoming query $Q$ against $S_{\text{ctx}}$ through the function $f_{\text{match}}$:

\[
f_{\text{match}}(Q, S_{\text{ctx}}) = 
\begin{cases}
    S_i, & \text{if } \exists S_i \in D_{\text{services}} \\
    Q_{\text{gen}}, & \text{otherwise}
\end{cases}
\]

where $Q_{\text{gen}}$ represents the refined service generation query which is achieved using RefineGenerationQuery function (line-14 of Algorithm \ref{alg:service_gen}) which would be passed to the Service Generator (line-15), another LLM agent with coding capabilities. This would contain instructions to write code, with relevant database information (their schema, location etc.) required to generate these services. 

\begin{table}[h]
\setlength{\tabcolsep}{3pt}
\caption{Algorithm-\ref{alg:service_gen} Components executed by LLM agents}
\label{tab:service-gen-components}
\centering
\begin{tabular}{ll}
\toprule
\textbf{Function} & \textbf{Description} \\
\midrule
RefineGenerationQuery & Examines D$_{schema}$ relevant to query Q to \\
& prepare LLM coding prompt Q$_{gen}$ \\ \hline
GenerateService & Processes Q$_{gen}$ to generate and deploy S$_{new}$ \\ \hline
UpdateServiceContext & Updates S$_{ctx}$ with S$_{new}$ and adds to D$_{services}$ \\
\bottomrule
\end{tabular}
\end{table}
For example, extending our previous scenario, if a user requests crowd monitoring at historical sites:

\[
S_{\text{new}} = \{
    \text{name}: \text{crowd\_monitor}, \text{ } \theta_c
\}
\]
\[
\text{where } \theta_c = \{
    \text{loc}: \text{hist\_site},
    \text{time}: \text{now}
\}
\]

If $S_{\text{new}} \notin D_{\text{services}}$ but $D_{\text{schema}}$ contains relevant data structures for the monitoring of crowd density via the sensors and endpoints of these databases, the Service Generator creates and integrates the new service:

\[
D_{\text{services}} = D_{\text{services}} \cup \{S_{\text{new}}\}
\]

Upon the generation of the new service $S_{new}$ an entry is added describing its function and the parameters it expects  (Table-\ref{tab:service-definitions} entry for crowd monitor service).

 After the successful service generation, UpdateServiceContext algorithm \ref{alg:service_gen} signals the Service Manager
 to incorporate the new service into its context state, triggering
 necessary updates to both Context Management and its
 service registry in the Knowledge Management. The newly generated service $S_{new}$ is further transmitted to the Builder component for building the website.

\subsection{\textbf{Intelligent User Interface Generator}}

The Intelligent User Interface Generator transforms the identified user goals into a functional web application through three primary components: Service Discovery, Builder, and Hosting. 

\[
G_{user} = \{SI_1, SI_2, ..., SI_n\}
\]

$G_{user}$ represents the set of services identified from the goal parser phase, where each $SI_i$ represents a service with its parameters and requirements, where each $\text{SI}_i$ represents an identified service. The Service Discovery component matches between $G_{user}$ and available services in the service registry. For each service $SI_i \in G_{user}$, it searches for matches $M_i \subseteq D_{services}$ by evaluating service descriptions and parameter compatibility:

\[
M = \bigcup_{i=1}^{n} M_i,\]\[ \quad M_i = \{ S_j \mid \text{match}(\text{description}_{\text{SI}_i}, \text{description}_{S_j}) \]
\[\land \text{params}_{\text{SI}_i} \subseteq \text{params}_{S_j} \}
\]

where $\text{match}()$ represents the semantic matching function which compares the service descriptions. All the matches are forwarded to the builder for application generation. When $M_i = \emptyset$ for any $SI_i$, indicating that no matching service exists, the system initiates service generation. The Backend Generation creates a new service $S_{new}$ using the service description and available data schemas and returns it to the Builder.

Builder implements UI generation process which employs a hierarchical rendering system $R$ with specific handlers for different data types, some of which are listed below:

\[
R(d) = \begin{cases}
    R_{\text{metric}}(d), & \text{if } d \in \mathbb{R} \\
    R_{\text{list}}(d), & \text{if } d \in \text{List} \\
    R_{\text{dict}}(d), & \text{if } d \in \text{Dict} \\
    R_{\text{text}}(d), & \text{otherwise}
\end{cases}
\]

where each renderer implements specific visualization logic for its data type (e.g., $R_{\text{metric}}$ for crowd density, $R_{\text{list}}$ for restaurant listings). The Builder Component generates the complete application by combining these renderers with a base template $T$, where $\oplus$ represents the composition operator that combines multiple rendered components:

\[
App = T(M \cup \{S_{new}\}) \oplus \bigoplus_{i=1}^{n} R(SI_i)
\]

The final application is hosted through the Hosting Component, which manages service endpoints and provides URL access to users.

\subsection{\textbf{Context Management}}
It hosts all the active services in the system. These services are built on top of the sensor data in Knowledge Management, which stores the data of the IoT environment (e.g., routes service, crowd monitoring, and air quality sensors). The configurations for these services are maintained in Knowledge Management, which is updated if the underlying data changes.

\subsection{\textbf{Knowledge Management}}
It is Responsible for storing environment and system-related data like the sensor data, user information like user preferences, location of user, configuration files, and a service registry. This registry includes a list of available services with brief descriptions of their functionality and parameters. Sensors periodically transmit data through the interoperable system, contributing to Knowledge Management's context for ongoing operations.

\section{Experiments and Results}

The evaluation focuses on assessing the proposed approach through the following research questions.

\begin{itemize}
    \item \textbf{RQ1: Effectiveness in Identifying Functionalities:}  
    How effective is the approach in identifying the correct set of functionalities corresponding to existing components ?
    
    \item \textbf{RQ2: Accuracy in Dynamic Service Generation:}  
    How accurate is the approach in dynamically generating services?  

    \item \textbf{RQ3: Effectiveness in System Generation:}  
    What is the effectiveness of the approach in generating the system as a whole?

    \item \textbf{RQ4: Efficiency in Application Generation:}  
    What is the efficiency of the approach in generating applications?
\end{itemize}
\subsection{Evaluation Setup} We evaluated our system using Hyderabad as a case study of a smart city implementation. The system comprises $9$ web-services: Air Quality, Crowd Monitoring, Event Notifier, Historic Information, Restaurant Locator, Travel Options, Water Quality, Exhibition Tracker, and Event Ticket Vendor. Several services operate on static contextual data feeds (e.g., Historic Information), while others process real-time data from a network of 12 distributed sensors. We implemented the IoT environment simulation using CupCarbon \footnotemark for realizing the architecture and generating sensor data based on domain-appropriate statistical distributions, with the core system in Python. we employed a two-fold evaluation strategy: (i) a multi-agent simulation framework modeling Tourist-Guide interactions across 100 experimental runs, and (ii) a user study ($n=15$) focusing on real-world usability and service adaptation quality.
\footnotetext{CupCarbon: \url{https://cupcarbon.com/}}
 \subsubsection{Tourist-Guide Simulations}
The system evaluation employed OpenAI's GPT-4o-mini \cite{openai2024gpt4technicalreport}, DeepSeek-V2.5 \cite{deepseekai2024deepseekv2strongeconomicalefficient} and CodeQwen1.5-7B \cite{bai2023qwentechnicalreport} models through the LangChain framework. We chose these based on the EvalPlus \cite{liu2023is} leader board. Experiments were conducted on an Nvidia L40S GPU with 8 vCPU, 62 GB RAM, and 48GB VRAM for hosting the CodeQwen1.5-7B model on HuggingFace. For GPT-4o-mini and DeepSeek-V2.5 model interactions, we utilized LangChain's OpenAI API. All these models were run with a temperature parameter of 0.7 based on preliminary experimentation.  
To evaluate the Goal Management, we designed a multi-agent simulation framework using CrewAI\footnotemark, modeling interactions between a Tourist agent and a Travel Guide agent (implementing our Goal Management's instruction set). The Tourist agent samples from $25$ predefined goals, generated through prompt engineering with domain-specific system knowledge, with time constraints uniformly distributed between $1-5$ hours. Each goal has an associated ground truth set of required services for validation.
\footnotetext{CrewAI: \url{https://www.crewai.com/}}
The goals were classified into concrete and ambiguous categories ($18:7$). Concrete goals have predictable mappings, such as ``Planning to visit Ramoji Film City'' mapping to \textit{ticket\_purchase} and \textit{travel\_options}. Ambiguous goals like ``First time in Hyderabad! Want to start with the locals' favorites'' may trigger multiple services (e.g., \textit{restaurant\_finder}, \textit{crowd\_monitor}, \textit{travel\_options}) based on conversation flow.
The simulation involved three sequential passes of Tourist-Guide interactions for service identification, repeated 100 times. While additional services could enhance user experience, we limit suggestions to avoid overwhelming users with options beyond their original goal.

\subsubsection{User Evaluation}
For complementing our simulation-based evaluation, we conducted a user study with students from the International Institute of Information Technology, Hyderabad (IIIT-H), which focused on understanding system effectiveness and overall user satisfaction through both quantitative metrics and qualitative feedback. The study involved participants ($n=15$) from diverse academic backgrounds within IIIT-H, specifically comprising 3 Ph.D. students (2 Computer Science and 1 Electronics/Communications Engineering), 5 Electronics/Communications Engineering students (B.Tech by M.S.), and 7 Computer Science students (B.Tech by M.S.). The participants were given a brief overview of the system's capabilities and were encouraged to interact with it based on their interests and needs. Each participant interacted with the system for approximately 10-15 minutes, with feedback collected through an integrated form in the user interface. The feedback mechanism collected three types of Quantitative Metrics: application rating, service accuracy rating, and service relevance rating (on a 1-5 likert scale), and Qualitative Feedback comprising query summaries, missing service identification, unnecessary service identification, and improvement suggestions. To further evaluate the effectiveness of dynamically generated services, we implemented a service rotation mechanism where three services were deliberately kept offline and replaced with generated implementations during each participant interaction, without informing participants, to assess integration seamlessness.

For assessing application generation efficiency, we integrated a metrics collection system with the Intelligent User Interface generator. The evaluation examined three critical metrics: total generation time (comprising dialogue latency, service discovery, template rendering, and deployment), token usage (aggregating input tokens from user queries, processing tokens from system context, and completion tokens from LLM responses), and build times per session. Given that service generation is not activated in every test scenario, we conducted a separate performance analysis of this component to ensure unbiased assessment.

\subsection{Results \& Discussions}
\noindent \textbf{RQ1: Effectiveness in Identifying Functionalities}

To evaluate the Goal Management's effectiveness, we analyze service identification accuracy using four key metrics defined in Table-\ref{tab:metrics-definition} in our simulation. The evaluation compares services identified after the third conversation pass against ground truth mappings derived from our tourism domain requirements.

\begin{table}[ht]
\setlength{\tabcolsep}{2pt}
\caption{Evaluation Metrics for Tourist-Guide simulation}
\label{tab:metrics-definition}
\centering
\begin{tabular}{l|p{7.5cm}}
\toprule
\textbf{Metric} & \textbf{Definition} \\
\midrule
Precision (P) & Ratio of correctly identified services to all identified services \\
\hline
Recall (R) & Ratio of correctly identified services to actual required services \\
\hline
F1 Score & Harmonic mean of precision and recall (2PR/(P+R)) \\
\hline
Parameter & Accuracy of identified service parameters (e.g., exact \\
Accuracy & locations, cuisines) against ground truth \\
\bottomrule
\end{tabular}
\end{table}

Analysis of the simulation results presented in Table-\ref{tab:goal-parser-categories} demonstrates comparable performance metrics between GPT-4o-mini and DeepSeek-V2.5 in service identification tasks. We noticed that both GPT-4o-mini and DeepSeek-V2.5 consistently respected time constraints while providing travel plans to the Tourist unlike CodeQwen1.5-7B which suggested plans spanning multiple days, exceeding the specified time constraints. CodeQwen1.5-7B exhibits lower precision values, displaying a tendency toward over-identification of required services. This over-identification introduces unnecessary complexity into the system architecture and imposes increased computational overhead during the build process.
\begin{table}[!htbp]
\setlength{\tabcolsep}{4pt}
\caption{Goal Parser Performance by Category}
\label{tab:goal-parser-categories}
\centering
\begin{tabular}{llcccc}
\toprule
\textbf{Model} & \textbf{Category} & \textbf{Precision} & \textbf{Recall} & \textbf{F1} & \textbf{Parameter} \\
& & & & & \textbf{Accuracy} \\
\midrule
CodeQwen1.5-7B & Ambiguous & 0.450 & 0.806 & 0.553 & 0.116 \\
& Concrete & 0.206 & 0.609 & 0.288 & 0.051 \\
& \textbf{Overall} & 0.282 & 0.670 & 0.370 & 0.071 \\
\midrule
GPT-4o-mini & Ambiguous & 0.683 & 0.795 & 0.730 & 0.549 \\
& Concrete & 0.467 & 0.773 & 0.559 & 0.739 \\
& \textbf{Overall} & 0.523 & 0.778 & 0.603 & 0.690 \\
\midrule
DeepSeek-V2.5 & Ambiguous & 0.681 & 0.788 & 0.725 & 0.585 \\
& Concrete & 0.492 & 0.830 & 0.591 & 0.743 \\
& \textbf{Overall} & 0.554 & 0.816 & 0.635 & 0.691 \\
\bottomrule
\end{tabular}
\end{table}

For the user evaluation (see Table-\ref{tab:user-satisfaction}) study, tourism-focused ($40\%$) and dining-related ($53\%$) queries dominated user sessions, with $67\%$ involving multi-service combinations. Restaurant Finder ($53\%$), Travel Options ($47\%$), and Historical Information ($33\%$) were the most frequently requested services. User feedback identified crowd monitoring (7 instances), air quality (3), and water quality (2) as desired additional services.
\begin{table}[ht]
\setlength{\tabcolsep}{4pt}
\caption{User Satisfaction Metrics}
\label{tab:user-satisfaction}
\centering
\begin{tabular}{lccc}
\toprule
\textbf{Metric} & \textbf{Average Rating (out of 5)} & & \\
\midrule
Application Rating & 4.0 & & \\
Accuracy Rating & 4.1 & & \\
Relevance Rating & 4.2 & & \\
\bottomrule
\end{tabular}
\end{table}
User studies highlighted the need for improved local data processing, particularly for proximity-based routing and recommendations. These insights suggest optimization areas aligning with our mixed-initiative vision, especially in collaborative monitoring and user-adaptive location services.

\smallskip
\noindent \textbf{RQ2: Accuracy in Dynamic Service Generation}

To evaluate the quality of dynamically generated services, we conducted multiple generation attempts (three per service) across our 9 services. We used CodeBERTScore \cite{zhou2023codebertscoreevaluatingcodegeneration} to assess the semantic similarity between generated and reference implementations, measuring four key aspects: precision (code correctness), recall (code completeness), F1-score (balanced measure), and F3-score (emphasizing on code completeness).

\begin{table}[ht]
\setlength{\tabcolsep}{4pt}
\caption{Service Generation Code Similarity}
\label{tab:code-similarity}
\centering
\begin{tabular}{lcccc}
\toprule
\textbf{Model} & \textbf{Precision} & \textbf{Recall} & \textbf{F1} & \textbf{F3} \\
\midrule
CodeQwen1.5-7B & 0.86 ± 0.02 & 0.79 ± 0.03 & 0.83 ± 0.02 & 0.80 ± 0.03 \\
DeepSeek-V2.5 & 0.91 ± 0.01 & 0.85 ± 0.03 & 0.88 ± 0.02 & 0.86 ± 0.03 \\
GPT-4o-mini & 0.90 ± 0.01 & 0.85 ± 0.03 & 0.87 ± 0.01 & 0.85 ± 0.02 \\
\bottomrule
\end{tabular}
\end{table}
As shown in Table \ref{tab:code-similarity}, DeepSeek-V2.5 achieved the highest overall performance with an F1-score of 0.88, outperforming CodeQwen1.5-7B by 6\% and comparable to GPT-4o-mini. Notably, all models maintained high precision (\(\ge\) 0.86), indicating reliable code generation quality. The relatively lower recall scores, particularly for CodeQwen1.5-7B (0.79), suggest occasional omissions in implementing complete functionality. These contrasting results from Table-\ref{tab:goal-parser-categories} suggest that while DeepSeek-V2.5 and GPT-4o-mini exhibit consistent performance across both service identification and code generation tasks, CodeQwen1.5-7B shows task-specific performance variations that could impact its suitability for general-purpose service generation in IoT environments. 

\smallskip
\noindent \textbf{RQ3: Effectiveness in System Generation}\\
We keep track of the total tokens required to generate these services across the models along with end-to-end latency
(including API request/response time) in Table-\ref{tab:service-generation}. \\
\begin{table}[!htb]
\setlength{\tabcolsep}{4pt}
\caption{Service Generation }
\label{tab:service-generation}
\centering
\begin{tabular}{lcc}
\toprule
\textbf{Model} & \textbf{Time (s)} & \textbf{Tokens} \\
\midrule
CodeQwen1.5-7B & 7.67 ± 0.26 & 3482.00 ± 19.80 \\
DeepSeek-V2.5 & 42.40 ± 4.52 & 4376.25 ± 228.17 \\
GPT-4o-mini & 25.66 ± 2.55 & 2063.17 ± 191.76 \\
\bottomrule
\end{tabular}
\end{table}
While GPT-4o-mini and DeepSeek-V2.5 achieved 100\% service generation success rate, CodeQwen1.5-7B only succeeded in 37\% of attempts. On inspection, we found that CodeQwen1.5-7B's performance limitations stemmed from (1) inconsistent instruction following and (2) JSON formatting errors.

\begin{figure}[!htb]
    \centering
    \includegraphics[width=0.8\linewidth]{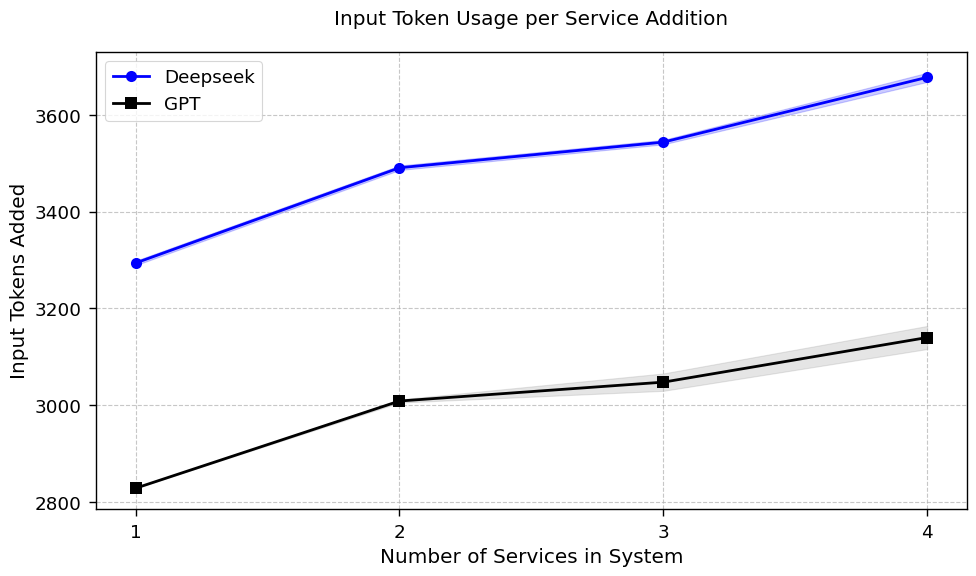}
    \caption{Token consumption scaling with increasing number of services for GPT-4o-mini and DeepSeek-V2.5. The x-axis represents the number of services and the y-axis shows the corresponding input token count.}
    \label{fig:token-scaling}
\end{figure}

To evaluate scalability in mixed-initiative contexts, we analyze how the input token consumption scales with an increasing number of services in the system. Figure~\ref{fig:token-scaling} illustrates this relationship across GPT and DeepSeek model, where the token usage pattern diverges significantly even for small number of services. This rise is primarily seen due to the addition of extra services leading to the Description Refiner requiring more tokens to process the system state received by the Service Manager.
While DeepSeek-V2.5 and GPT-4o-mini have comparable pricing (0.14 USD and 0.15 USD per 1M input tokens)\footnotemark, their actual costs differ due to variations in token consumption, directly impacting adaptive service generation costs. GPT-4o-mini, with its relatively compact architecture and lower token usage (Table-\ref{tab:service-generation}), demonstrates more efficient performance for dynamic interactions compared to DeepSeek-V2.5's 238 billion parameter architecture.
\footnotetext{As of 2024-12-10 on billing websites}
\smallskip

\noindent \textbf{Results for RQ4: Efficiency in Application Generation}
We evaluated the system using 15 scenarios from our human evaluation study.\subsection{Performance Analysis}
The system achieved an average total generation time of 23.10 ± 6.47 seconds. Table~\ref{tab:generation-breakdown} presents the detailed breakdown of the processing stages.
\begin{table}[ht]
\setlength{\tabcolsep}{4pt}
\caption{Generation Process Time Breakdown}
\label{tab:generation-breakdown}
\centering
\begin{tabular}{lcc}
\toprule
\textbf{Processing Stage} & \textbf{Mean (s)} & \textbf{SD (s)} \\
\midrule
Multi-pass conversation processing & 18.25 & 5.12 \\
Service identification \& parameter extraction & 3.82 & 1.14 \\
Template rendering \& application assembly & 1.03 & 0.31 \\
Final deployment & 0.004 & 0.002 \\
\bottomrule
\end{tabular}
\end{table}

Analysis of token distribution is presented in Table~\ref{tab:token-distribution}.

\begin{table}[ht]
\setlength{\tabcolsep}{4pt}
\caption{Token Usage Distribution}
\label{tab:token-distribution}
\centering
\begin{tabular}{lccc}
\toprule
\textbf{Token Type} & \textbf{Count (Mean ± SD)} & \textbf{\% of Total} \\
\midrule
Input tokens & 101.8 ± 70.12 & 1.25\% \\
Processing tokens & 7,308.1 ± 2,607.49 & 89.51\% \\
Completion tokens & 755.0 ± 281.28 & 9.24\% \\
\bottomrule
\end{tabular}
\end{table}

The system demonstrated decent build performance, with an average build time of 4.85 ± 1.98 milliseconds. This sub-10 ms build time was anticipated, as the builder only needs to render the application by sending it to the hosting component. Table~\ref{tab:app-generation} summarizes the overall performance metrics.

\begin{table}[ht]
\setlength{\tabcolsep}{4pt}
\caption{Application Generation Performance Metrics}
\label{tab:app-generation}
\centering
\begin{tabular}{lccc}
\toprule
\textbf{Metric} & \textbf{Mean ± SD} & \textbf{Min} & \textbf{Max} \\
\midrule
Total Duration (s) & 23.10 ± 6.47 & 13.46 & 33.08 \\
Total Token Usage & 8164.90 ± 2718.89 & 5531 & 13991 \\
Build Time (ms) & 4.85 ± 1.98 & 3.50 & 10.49 \\
\bottomrule
\end{tabular}
\end{table}

\begin{figure}[!htb]
    \centering
    \includegraphics[width=0.8\linewidth]{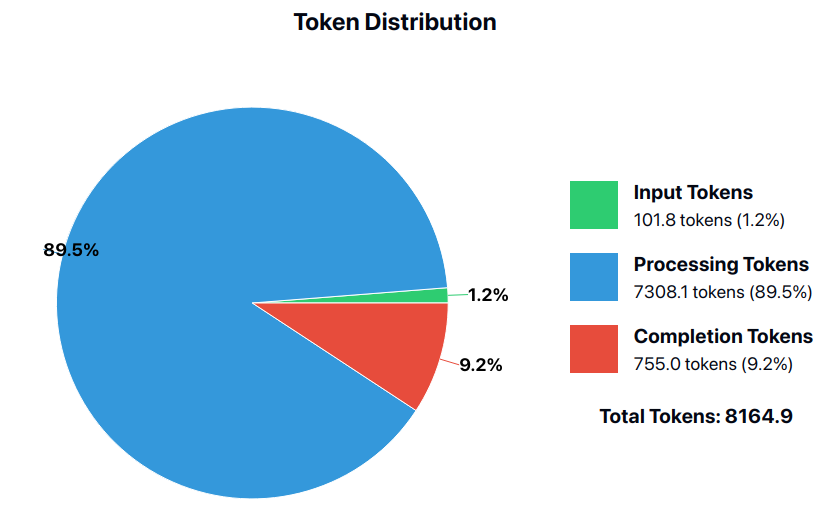}
    \caption{Token distribution during application generation. The high proportion of processing tokens (89.51\%) indicates potential for optimization through context management improvements.}
    \label{fig:token-distribution}
\end{figure}

\subsection{Service Generation Analysis}

The Backend generation component was evaluated by generating each of our nine services ten times. The evaluation revealed consistent results across different service types, with an average generation time of 15.53 seconds. The metrics are summarized in  Table~\ref{tab:service-metrics}.

\begin{table}[ht]
\setlength{\tabcolsep}{4pt}
\caption{Service Generation Performance Metrics}
\label{tab:service-metrics}
\centering
\begin{tabular}{lc}
\toprule
\textbf{Metric} & \textbf{Value} \\
\midrule
Average processing time (s) & $15.53 \pm 1.12$ \\
Total token usage & $4,992.89 \pm 180.29$ \\
Coefficient of Variation (\%) & $3.61$ \\
\bottomrule
\end{tabular}
\end{table}

The Coefficient of Variation (CV), calculated as (Standard Deviation / Mean) × 100, measures dispersion across different metrics. A CV of 3.61\% indicates high consistency in generating any service regardless of its type. When service generation is incorporated into the total application generation metrics, we observe a total duration of 38.63 seconds (23.10 ± 6.47 + 15.53 ± 1.12) and total token usage of 13,157.79 tokens (8,164.90 ± 2,718.89 + 4,992.89 ± 180.29).

\renewcommand{\thefootnote}{} % Suppress footnote numbering
\footnotetext{Code available on GitHub: \url{https://github.com/sa4s-serc/SAS_llm_query/tree/iot-prototype}}
\renewcommand{\thefootnote}{\arabic{footnote}} % Restore footnote numbering

\section{Threats to Validity}

\textbf{External Validity:} The Goal Management evaluation faces generalization constraints with a limited dataset of 25 predefined goals in the tourism domain. The implementation's focus on Hyderabad's $9$ specific services may restrict generalizability to cities with different infrastructure requirements. While all the of participants showed willingness for future use, the student-only participant pool and short interaction duration (10-15 minutes) limit comprehensive understanding of long-term usage patterns.

\textbf{Internal Validity:} The uniform temperature setting (0.7) across models might not represent optimal individual configurations. CodeBERTScore, justified by standardized service generation, may not fully capture semantic code differences. The fixed three-pass conversation system could potentially miss interaction patterns affecting service discovery accuracy. The service rotation mechanism provides insights but may not represent all production environment failure modes.

\textbf{Construction Validity:} Our service identification approach uses precision and recall metrics against predefined mappings, which may not fully capture user preference variations. The metrics might not account for additional beneficial services or context-specific requirements. While the human evaluation study \((n=15)\) provides real-world validation, its small sample size limits generalizability.

\section{Related Work}

Recent advances in adaptive systems and human-centric computing have produced notable approaches for managing IoT environments. Mayer et al. \cite{7444198} propose a semantic technology-based framework for user goal modeling, while Abughazala et al. \cite{abughazala2021human} combine Agent-Based Modeling with architectural techniques to translate human interactions into actionable configurations. De Sanctis et al. \cite{de2021user} developed a multi-level self-adaptation approach driven by users at run-time. Moghaddam et al. \cite{moghaddam2023user} introduce an emotion-aware framework optimizing QoE and QoS through Model-Free Reinforcement Learning, and Yigitbas et al. \cite{yigitbas2020integrated} contribute a model-driven development approach using domain-specific languages for context modeling. 
Qian et al.\cite{7479556} proposed MobiGoal, a framework using a runtime goal model to adaptively schedule user goals and execute tasks. While MobiGoal defines goals during design-time to guide application development and runtime execution, our approach allows user goals to emerge dynamically at runtime without requiring users to understand the goal model. Recent advances in Large Language Models (LLMs) have opened new possibilities for bridging human intent and system adaptation. While traditional program synthesis focused on predefined approaches \cite{DBLP:journals/corr/RaghothamanU14}, modern LLMs like GPT-4 \cite{jiang2024selfplanningcodegenerationlarge}\cite{he2024doespromptformattingimpact} and DeepSeek-V2 \cite{deepseekai2024deepseekv2strongeconomicalefficient} demonstrate proficiency in understanding user intent and generating functional code, enabling systems to dynamically interpret and adapt to user goals through natural language interaction.
Further, significant progress in multi-agent \cite{wu2023autogenenablingnextgenllm} work which provide mechanisms for agents to share context, negotiate goals, and collectively adapt to changing environments.
While existing human-centric IoT systems often rely on predefined adaptation mechanisms, our work introduces a novel architecture integrating large language models (LLMs) to enable real-time service adaptation through natural language interactions. Inspired by Spalazzese et al.'s \cite{spalazzese2023shaping} Mixed-Initiative paradigm of collaborative goal refinement, we extend IoT system capabilities by incorporating LLM-based components for dynamic runtime adaptation. By combining natural language processing, sensor data integration, and runtime adaptability, our approach advances flexible, user-centric solutions for evolving IoT environments.

\section{Conclusion \& Future Directions}

In this work we advance the IoT-Together paradigm by presenting a goal-driven system utilizing Large Language Models for user-system collaborative applications which can be dynamically composed at runtime. Our system highlights: (1)~a multi-pass dialogue framework translating natural language goals into system requirements, (2)~an LLM-based service generation pipeline with code generation capabilities, and (3)~a template-based interface generation system dynamically adapting to diverse service combinations. Our evaluation in a smart city case study demonstrates the potential of LLMs in enabling mixed-initiative interaction and automated service generation through intelligent user-system collaboration.

Our proposed approach establishes foundation for advancing mixed-initiative IoT architectures that can dynamically adapt while maintaining operational reliability in complex deployment scenarios. While LLM-based components introduce performance variability, deterministic components maintain system stability. Future research directions include exploring deterministic and probabilistic component combinations to improve user-system collaboration, with concrete functional testing to improve reliability. This involves developing LLM-based testing frameworks to validate generated services before integration, requiring testable code generation patterns and enhanced builder capabilities. Given the substantial computational resources of large language models, future work could explore dynamic model selection based on energy constraints, and investigate lightweight alternatives for less complex tasks within the system.

\bibliographystyle{ieeetr}
\bibliography{references}

\end{document}